\documentclass[10pt, conference, compsocconf]{IEEEtran}

\usepackage[utf8]{inputenc}
\usepackage{amsmath}
\usepackage[english]{babel}
\usepackage{graphicx}
\usepackage{multicol}
\usepackage{subfigure}
\usepackage[numbers,sort&compress]{natbib}
\usepackage{color}

\newcommand\addedstuff[1]{#1}

\begin{document}

\title{A Cognitive-inspired Model for Self-organizing Networks}

\author{
\IEEEauthorblockN{
	Daniel Borkmann\IEEEauthorrefmark{1},
	Andrea Guazzini\IEEEauthorrefmark{2}\IEEEauthorrefmark{3},
	Emanuele Massaro\IEEEauthorrefmark{4}, 
	Stefan Rudolph\IEEEauthorrefmark{6}}

\IEEEauthorblockA{
	\IEEEauthorrefmark{1}Communication Systems Group, 
	ETH Zurich, Switzerland}
%	Email: daniel.borkmann@tik.ee.ethz.ch}

\IEEEauthorblockA{\IEEEauthorrefmark{2} Institute for Informatics and Telematics, 
	National Research Council, Pisa, Italy}
%	Email: andrea.guazzini@iit.cnr.it}

\IEEEauthorblockA{\IEEEauthorrefmark{3} Department of Psychology,
	University of Florence, Italy}

\IEEEauthorblockA{\IEEEauthorrefmark{4}Department of Informatics and Systems, 
	University of Florence, Italy}
%	Email: emanuele.massaro@unifi.it}

\IEEEauthorblockA{\IEEEauthorrefmark{6} Organic Computing Group,  
	University of Augsburg, Germany}
%	Email: stefan.rudolph@informatik.uni-augsburg.de}
}

\maketitle

\begin{abstract}
In this work we propose a computational scheme inspired by the workings of
human cognition. We embed some fundamental aspects of the human cognitive system
into this scheme in order to obtain a minimization of computational resources
and the evolution of a dynamic knowledge network over time, and apply it to computer
networks.
Such algorithm is capable of generating suitable strategies to explore huge graphs
like the Internet that are too large and too dynamic to be ever perfectly known.
The developed algorithm equips each node with a local information about possible
hubs which are present in its environment. Such information can be used by a node
to change its connections whenever its fitness is not satisfying some given requirements.
Eventually, we compare our algorithm with a randomized approach within an
ecological scenario for the ICT domain, where a network of nodes
carries a certain set of objects, and each node retrieves a subset at a certain time,
constrained with limited resources in terms of energy and bandwidth. We
show that a cognitive-inspired approach improves the overall networks topology
better than a randomized algorithm.
\end{abstract}


\IEEEpeerreviewmaketitle

\section{Introduction}
%Among the capability of the human cognitive system that is attracting most of
%the computer scientists, there is the ability of humans to develop local
%algorithms able to exploit what might be called the ``collective human computation''.

\addedstuff{Among the capabilities of the human cognitive system that is attracting most 
computer scientists, there is the ability of humans to develop local algorithms
able to exploit what might be called ``collective human computation''. As
collective human computation, we refer to the natural synchronization between
the cognitive elaborations made by a person which is immersed into group
dynamics. In such a condition, human beings analyze only some relevant information
coming from the group, giving to the group only some relevant contributions for
the general problem which is faced. In this way, the group can be described as
more than just the sum of its single components \cite{Hoffman196699}.}

When being faced with insufficient data or insufficient time for rational
processing, humans have developed strategies that allow to take decisions
in these situations. 
In general, such an effect has been well described in the cognitive heuristics program
proposed by Goldstein and Gigerenzer, which suggest starting from
fundamental psychological mechanisms in order to
design models of heuristics~\cite{Gigerenzer2002}.

Only some relevant information are extracted from the environment while the rest
is interpolated for building our knowledge.
Among others, Milgram et al. have shown experimentally how humans are able to
adopt effective strategies to solve very complex problems, exploiting optimally
their partial knowledge of their environment~\cite{Dunbar1, milgram1967smallWorld}.

This kind of human distributed computing has been studied deeply only from the
perspective of disciplines such as social cognition and social psychology, while it
is not yet well known in other domains.

The social cognition domain studies human cognition as characterized by the use
of ``fast and frugal'' solutions, that are specialized for a social context in which
we live using a bounded rationality
and limited computational resources~\cite{Simon55}.
Therefore, the aim of our work is to assemble a working computational scheme
inspired by the operating principles of human cognition, based on general
assumptions about cognitive high-level functions.
%Fundamentally, we implemented a computational scheme that mimics what is
%known about the way humans take decisions ~\cite{Simon55}.
% embedding both,
%bounded rationality instances and the concept of satisfaction.
% instead of the optimizing approach
%to the knowledge representation~\cite{Simon55}.

%This approach promises to embed some fundamental
%aspects of the human cognitive system in a computational model, such as the
%minimization of the computational resources needed for the task, and the evolution of a dynamic
%knowledge network capable of generating suitable strategies to networks like
%the World Wide Web, e.g. which is too large and too dynamic to ever be
%fully/perfectly known~\cite{Massaro2012}.

\addedstuff{This approach promises to embed some fundamental aspects of the human
cognitive system in a computational model in order to obtain a minimization of 
computational resources needed for the task and the evolution of a dynamic
knowledge network capable of generating strategies suitable for networks
like the Internet, which are too large and too dynamic to ever be
fully/perfectly known~\cite{Massaro2012}.}

The fundamental aspects on which we focused our modeling, involves the
spread of information through a human network, and the knowledge representation
arising from the dynamics of short-term memory (STM) and long-term memory (LTM).
The passage of information between STM and LTM occurs through a simple
cognitive heuristic approach, which compatibly with their computational capacity
reduces the dimensionality of the information required to represent the
environment in a dynamic manner.

%In previous work, we applied such an approach to the community detection
%problem, which can be considered as a task of great importance in many
%disciplines~\cite{Waaserman,Scott,Mendes,Strogatz,Albert}, where systems can be
%represented as graphs. Clearly, community detection can be easily linked to clustering of data
%and consequently to the management of resources (e.g. bandwidth and contents
%management).

\addedstuff{In our previous work~\cite{Massaro2012}, we applied such an approach
to the community detection problem, which can be considered as a task of great
importance in many disciplines~\cite{Waaserman,Scott,Mendes,Strogatz,Albert},
where systems can be represented as graphs. The first version of the algorithm
was characterized by a two step procedure (e.g. discovering and elaboration
phases), in which the effect of the nodes' connectivity on the information
spreading was exploited by nodes to assess a first approximation of the topology
of the network. In this work, we present a second version of the algorithm
in which the third phase was added in order to refine the topology detection by
a cognitive inspired strategy which embeds the cognitive dissonance theory
\cite{citeulike:235567}. 
%Moreover, the upgrading of the algorithm makes the model
%independent from parameters (i.e. $m$ and $\alpha$). Thus, making them dynamic
%and adapting dependent on the ``knowledge'' of the nodes.
}

In general, as the Internet nowadays, human social networks have to be
considered as a continuum of nested communities whose boundaries are
somewhat arbitrary~\cite{Lanc}.

Here, we propose such a tool for detecting communities in complex networks
using a local algorithm, applied as a cellular automaton. In this approximation,
a node is just modeled as a memory and a set of links to other nodes. The
information about neighbouring nodes is propagated using a standard diffusion
process, and elaborated locally using a non-linear competition process among
the information. This process can be considered an implementation of the
``take the best'' heuristic~\cite{Gigerenzer2011}, which relies on the
assumption that the most relevant or easily detectable information gives an
accurate estimate of the
frequency of the related event/contents in the population. The result of the
algorithm equips each node with information about possible hubs or
super-nodes present in its environment, and such information can be used by
the node to rewire its connections whenever its fitness does not satisfy some
given requirements.

\addedstuff{In real-world applications, such a process can be engineered
within the ICT domain. Consider for instance resilience and
scalability effects in service ecosystems. There, one important factor
is to decentralize services. This can be done with the help of creating overlay networks on
top of large-scale ones such as the Internet.
An adaptive, intelligent or even resource-optimizing algorithm plays a crucial
role for the (self-)maintenance of such systems.}

In that way, we could tackle the first steps to create an intelligent,
semi-structured peer-to-peer overlay network from an unstructured one, e.g. like a
\textit{self-optimizing} FastTrack \cite{Liang:2006:FOM:1141103.1141110} network. FastTrack itself
uses a semi-structured overlay network with a mix of \textit{designated} super-nodes
and normal nodes.
The latter have to connect to one of the super-nodes in order to minimize
redundant communication overhead. There, participating nodes could retrieve
content at a certain time with given resource constraints (e.g. bandwidth,
energy, latency), detect super-nodes automatically during an operation, and
thus change their connections (and therefore the overlay topology) for better
conditions.

\addedstuff{The rest of this paper is structured as follows: section 2 describes the scenario
and section 3 the cognitive-inspired algorithm. In section 4, we evaluate our
cognitive-inspired algorithm with a randomized algorithm. Eventually, in section 5
we conclude our findings.}

%As a possible scenario let us consider an Internet node who wants to get a
%certain content at a certain time, given a limited resources (e.g. time,
%energy and bandwidth). For such a node is impossible to calculate the
%relevant network characteristics, besides it could use an individual strategy
%to discover if through its first neighbours it is able to gather the content
%it is searching for, determining on this bases if it belongs to the right
%community.

\section{Scenario}
\label{sec:sce}
As a first step towards such a real-world self-optimizing peer-to-peer
network, we consider the following \textit{simplified} approach: given $N$
individuals (nodes), labeled from $1$ to $N$, where each individual hosts exactly one
item. There are $I$ items distributed over individuals. Also, we label items
from $1$ to $I$, where $I \leq N$, thus two or more individuals can host 
same items. Each individual has a pre-defined maximum number of links, where
it can connect to other nodes. As a simplification, we can denote a link between
two individuals as ``wired''. During the initial state, not all links are wired.
Hence, some individuals still have free capacities in our network topology.
%Those free link capacities are Gaussian distributed among $N$.
For each node from $1$ to $N$, the free link capacities are uniformly distributed
within a given interval $[a,b]$.
Now, each individual acts greedy and wants to collect up to a maximum number
of unique items $I_{curr} \leq I_{max}$ from other individuals, where
$I_{max} \leq I$ and $I_{curr}$ defines the actual number of items that have been collected.
However, in collecting, an individual is constrained by a given budget/energy $C_{curr}=C_{max}$ it can
spend.
%While exploring its $i$th-degree neighbors ($i \ge 0$), it has to pay
%for the number of ``hops'' if it has enough budget left, so
%$C_{curr}\leftarrow{}C_{curr}-(1+i)$ (the ``payment'' itself costs $1$ energy point).
While exploring its $i$th-degree neighbors ($i>0$), it has to pay
for the number of hops if it has enough budget left, so that
$C_{curr}\leftarrow{}C_{curr}-i$.
Note that an individual does not have a global knowledge of the topology. After
this process has been completed by each node, a given fitness function $f$ can
be calculated. $f$ is then used in order to find ``weak nodes''. Candidates 
must give up one of their links and create a new one to a more ``promising''
node, e.g. to a super-node/hub. Hence, in each round of this process,
the topology will be partially changed by a set of weak nodes and a minor
randomly selected component of the system.

In this paper, we are evaluating our network's behaviour from two perspectives:
(i) $f_1:$ maximizing a node's $I_{curr}$, that is, each node shall collect as
many unique items as possible, so that $I_{max}-I_{curr}\rightarrow{}0$ while
complying to its energy constraint, (ii) $f_2:$ minimizing a node's
$C_{max}-C_{curr}\rightarrow{}0$ while having $I_{curr}=I_{max}$ items
collected. In this case, the pre-defined energy $C_{max}$ is sufficiently large
(no energy constraint) to collect $I_{max}$ items, so that the
system's focus is to minimize its overall energy.
In each round in (i) and (ii), $I_{curr}$ and $C_{curr}$ are reset and $f$
reevaluated.
%Thus, we're going to optimize the topology of our content distribution
%network regarding bandwidth costs and energy efficiency.
%Therefore,
We claim that by carefully choosing weak nodes and promising nodes for
rewiring links, we can optimize $f$ over time. 
%Hence, instead of just randomly
%selecting, we give each individual a bounded memory that provides knowledge
%about its surrounding for a better decision making as provided in the next
%section.
Hence, instead of just randomly selecting individuals, we give each of them a
bounded memory that provides knowledge about its surrounding for a better decision
making as provided in the next section. Thus, we make nodes self-aware of their
own ``world''.

%%% Why method? I would say that you can name it a math. model
\section{Algorithm}
\label{sec:model}
%Our model for the optimization of the network topology with respect to the
%given scenario consists of three parts:
%(i) a ``game'' phase, (ii) a discovery phase and (iii) a rewiring phase.
%
%In the first part, each node epidemically traverses their neighbors to collect
%unique items under a constraint of an upper energy limit. Depending on their
%available energy, they successively visit their first-degree neighbors, then
%their second-degree neighbors and so forth until no energy is left to ``pay''
%for an item:
%
%put equation here
%
%%\comment{Maybe we make the fitness/energy calculation more generic and explain
%%it separately for the 2 scenarios in the evaluation part?}
%
%Say sth more here, bla bla....
%
%The second part represents a cognitive exploration in which nodes become aware
%of their surrounding world. We have already observed that this kind of
%algorithm are useful also for detection communities in
%networks~\cite{Massaro2012, Bagnoli2012}. In our case, this cognitive part is
%being used to find potential ``hubs'' respectively ``non-hubs'' for each node,
%information that is being exploited in the third part, the ``rewiring phase''.
%
%\subsection{Game phase}
%\label{sec:mod1}
%ToDo

%\subsection{Discovery phase}
%\label{sec:mod2}
We create a local algorithm where an individual is simply modeled as a memory
and a set of connections to other individuals. The ``learning'' (nonlinear)
phase is modeled after competitions found in the chemical/ecological world, where
resources compete against each other in order to not fall into oblivion.
%phase is modeled after competitions found in the chemical/ecological world, where
%resources competing against each other in order to not falling into oblivion. 

%%% XXX move this to game phase
%As stated, we consider $N$ individuals, labeled from $1$ to $N$. 
We consider an unweighted undirected network with the adjacency matrix $A$:
the adjacency matrix of a finite graph $G$ on $n$ vertices is the $n \times n$
matrix where the non-diagonal entry $A_{ij}$ represents the presence $(A_{ij}=1)$
or the absence $(A_{ij}=0)$ of a link between the vertices $i$ and $j$.

Then each vertex $i$ in the graph is characterized by a state vector $S_i$
that represents its knowledge of the others. In our model, we consider $S$ as
a probability distribution, in particular $S_i^{(k)}$ is the probability that
individual $i$ belongs to the community $k$. 

Then $S_i^{(k)}$ is normalized using the index $k$. Considering $S=S(t)$ the
state matrix of the network at time $t$, with $S_{ik} = S_i^{(k)}$.  At
time $t = 0$ each node only knows about itself so $S_{ij}(0) = 1$ if $i=j$ and $0$
otherwise. As mentioned before, the competition phase is modeled analogous to a
chemical/ecological concept. Our algorithm is inspired by the concept of
\emph{diffusion and competitive interaction} in network structure introduced
by Nicosia et al.~\cite{Nicosia09}.

If two populations $x$ and $y$ are in competition for a given resource, their
total abundance is limited~\cite{murray2002}.  After normalization, we can
assume that $x+y=1$, i.e., where $x$ and $y$ are the frequency of the two species, and
$y=1-x$. The reproductive step is given by $x' = f(x)$, which we assume to be
represented by a power $x'=x^\alpha$. For instance, $\alpha=2$ models the birth
of individuals of a new generation after binary encounters of individuals
belonging to the old generation, with non-overlapping generations
(eggs laying)~\cite{frei}.

After normalization we obtain:
\begin{equation}
 x' = \frac{x^\alpha}{x^\alpha + y^\alpha} = \frac{x^\alpha}{x^\alpha + (1-x)^\alpha}.
\end{equation}
Introducing $z=(1/x)-1$ ($0\le z< \infty$), we get the map
\begin{equation}  
\label{ref:second}
 z(t+1) = z^\alpha(t), 
\end{equation}
whose fixed points (for $\alpha > 1$) are $0$ and $\infty$ (stable attractors)
and $1$ (unstable), which separates the basins of the two attractors. Thus, the
initial value of $x$, $x_0$, determines the asymptotic value, for
$0\le x < 1/2$, $x(t\rightarrow\infty) = 0$, and for
$1/2< x < 1$, $x(t\rightarrow\infty) = 1$.

%XXX
%\begin{figure*}[t!]
%\centering
%\subfigure[]
%{\includegraphics[width=11cm]{out.pdf}}
%\hspace{1mm}
%\subfigure[]
%{\includegraphics[width=6cm]{conn.pdf}}
%\caption{\label{fig:g1} (a) Simple \emph{toy network} network composed by nine
%nodes and two communities. (b) The cumulative distribution $P^{(S)}$ (dashed
%line, $P_j^{(S)}=\sum_i S_{ij}$, multiplied by five) and $P^{(A)}$ (solid 
%line, $P_j^{(A)}=\sum_i A_{ij}$, connectivity). The information propagation
%algorithm identifies communities by hubs (nodes $4$ and $6$ with higher
%connectivity) with $m=0.3$ and $\alpha = 1.4$.}
%\end{figure*}

% the dynamics defining
% can be usefulto define a prob distribution P_i, where after normalization x_ij 
%In order to define a probability distribution $P_i$ for a large number of components,
%it is necessary to normalize these values and so the dynamics becomes:
%If we extend the dynamics to a larger number of components, for a probability
%distribution $P_i$, the competition dynamics becomes
%\begin{equation}
%  P_{i}' = \frac{P_{i}^\alpha}{\sum_j P_{j}^\alpha}.
%  \label{ref:competizione}
%\end{equation}

%Then, 
The dynamics of the network are given by an alternation of communication
and elaboration phases.
In the communication phase, there is a diffusion of
information in which each node has a memory factor $m$; in this way, in
each time step nodes update the previous information with new information. Due to
this parameter, we can introduce some limitations into the algorithm as in the
human cognitive system such as the mechanism of oblivion and the timing
effects: the most recent information has more relevance than previous information ~\cite{Tulving82,Forster84}. 

We assume that nodes talk with each other and we suppose that nodes with high
connectivity degree have greater influence in the process of information's
diffusion. This is due to the fact that during a conversation it is more likely
to know a vertex with high degree instead of one that has few links. For
this reason, the information dynamics is a function of the adjacency matrix $A$.

Then, in the communication phase, the state of the system evolves as
\begin{equation}
  S\left(t+\frac{1}{2}\right) = m S(t) + (1-m) A S(t).
\end{equation}

%As described in the Eq. \ref{ref:competizione}, 
The competition phase is
modeled analogously to a competitive interaction between the nodes in the
network~\cite{Nicosia09}. In this way the dynamic of the model is given by the sequence
$S(t)\rightarrow S(t+\frac{1}{2})\rightarrow S(t+1)$:
\begin{equation}
\label{ref:bagnongen} 
 \begin{split}
   {S}_{ik}\left( t+\frac{1}{2}\right) &= m S_{ik}(t) + (1-m) \sum_j A_{ij} S_{jk}(t), \\ 
   S_{ik}(t+1) &= \frac{{{S}}_{ik}^{\alpha}(t+\frac{1}{2})}{\sum_j {{S}}_{ij}^{\alpha}(t+\frac{1}{2})}.
 \end{split}
\end{equation}

\begin{figure*}[t!]
\centering
\subfigure[]
{\includegraphics[width=3.5cm]{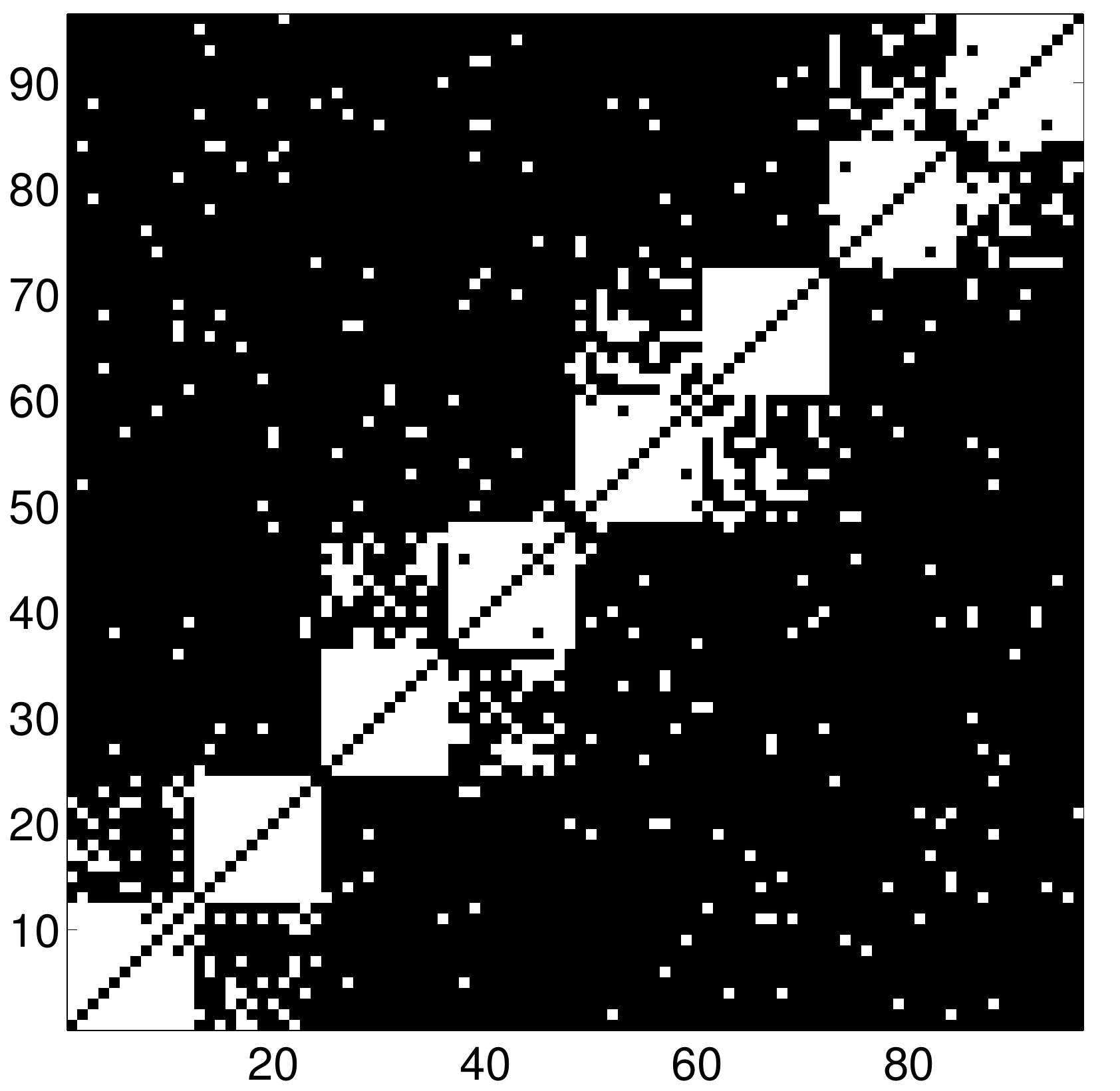}}
\hspace{.1mm}
\subfigure[]
{\includegraphics[width=3.5cm]{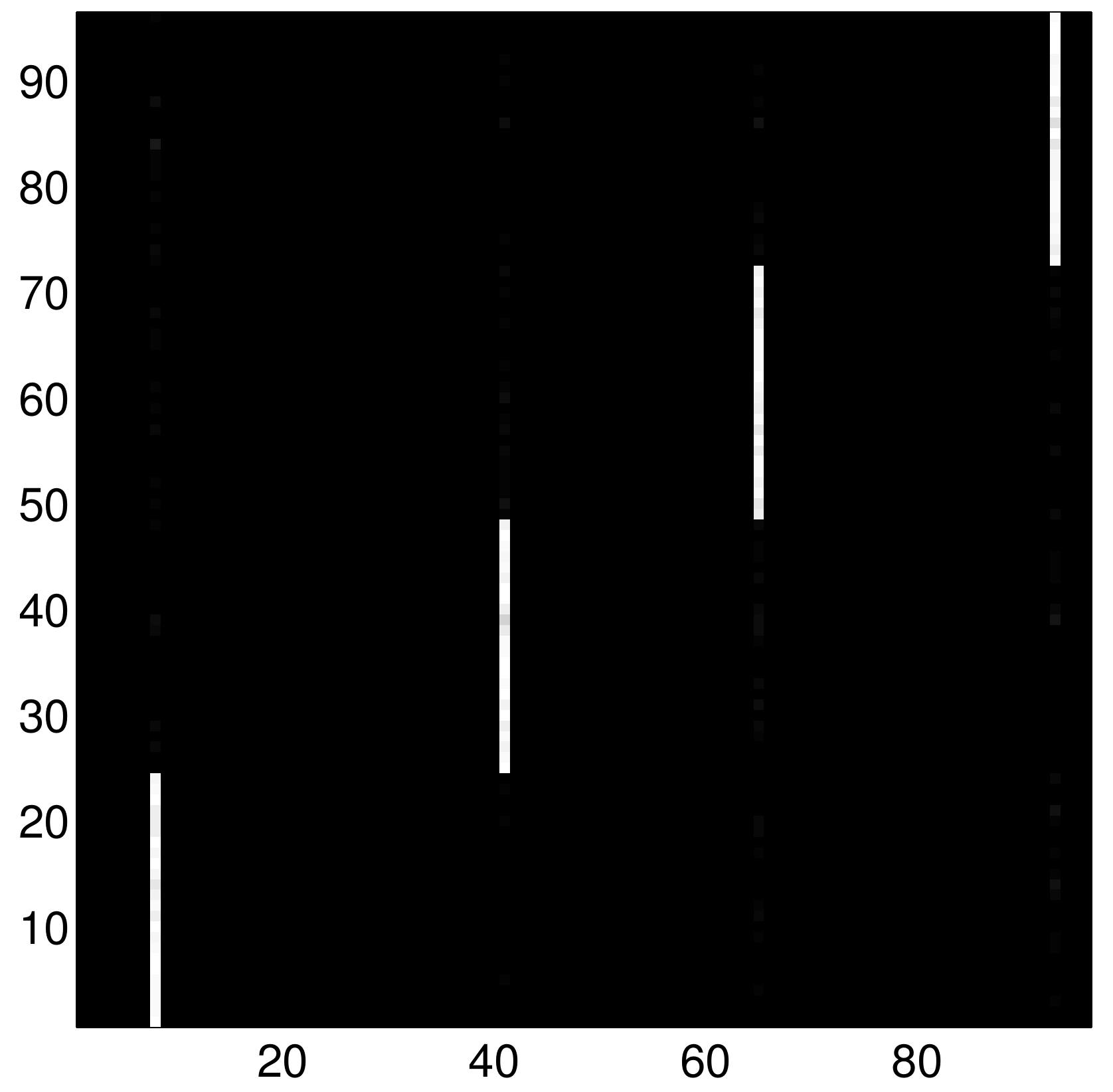}}
\hspace{.1mm}
\subfigure[]
{\includegraphics[width=3.5 cm]{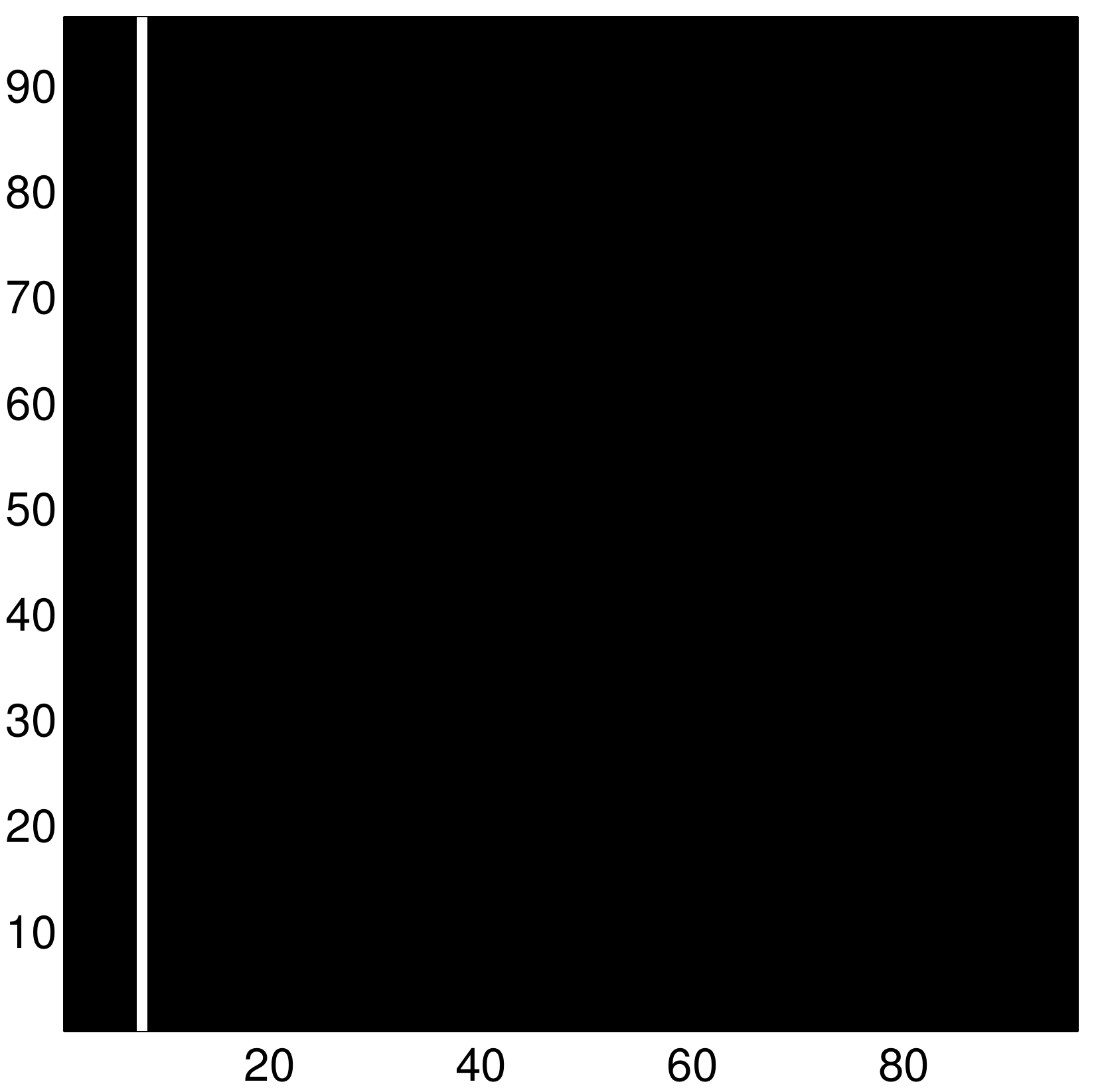}}
\hspace{.1mm}
\subfigure[]
{\includegraphics[height=3.7 cm]{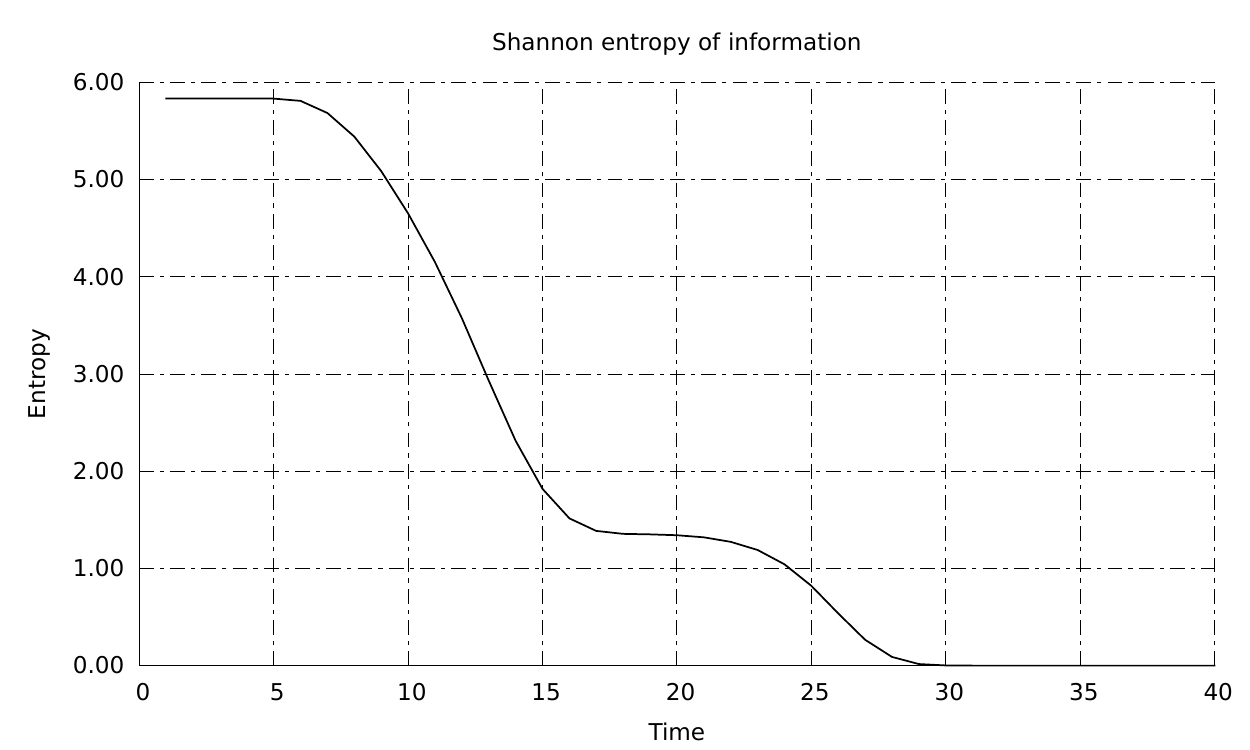}}
\caption{\label{fig:g2} (a) Hierarchical three-level network with $4$ principal
communities. (b) Final configuration of state matrix $S$ with $m = 0.7$ and
$\alpha = 1.4$. (c) Final configuration of state matrix $S$ with $m = 0.27$
and $\alpha = 1.25$: the final mono-cluster is identified by the major hub
in the network. (d) Entropy of information for the whole network during time
regarding the case (c).}
\end{figure*}

The node's memory is assumed to be large enough to contain all information about
other nodes, and the model is characterized by two free parameters: the
memory $m$ and the coefficient $\alpha$.
%\figurename~\ref{fig:g1} shows a
%result of the algorithm in a simple \emph{toy network} with fixed parameters.
As \figurename~\ref{fig:g2} shows, this model is correlated to
the values of parameters, and it is able to discover different final
structures and results. In \figurename~\ref{fig:g2}~(a), an example of a
hierarchical network in form of an adjacency matrix $A$ is represented, where a
three-levelled matrix is composed by $4$ blocks of $2$ sub-communities of $8$
nodes each, with a link probability that is respectively of $0.98$ inside
sub-community, $0.3$  in the first level of nested blocks, and $0.03$ among
blocks.  The white points indicate the presence of a link between the
node $i$ and the node $j$, $A_{ij} = 1 $. 
In \figurename~\ref{fig:g2} (b), the asymptotic configuration of the
matrix $S$ using $m = 0.7$ and $\alpha = 1.4$ is shown, while in
\figurename~\ref{fig:g2} (c) with $m = 0.27$ and $\alpha = 1.25$.
Finally, in \figurename~\ref{fig:g2} (d), the dynamic evolution of the entropy
of information, $E$ corresponding to the case (c), defined
as $E^{(S)} = - \sum_i P_i^{(S)} \log(P_i^{(S)})$, is represented, where $P_i^{(S)} = \sum_i S_{ij}$. 
%The entropy $E$ reaches the maximum only for the flat distribution,
%where each node knows only itself, and reaches a minimum (zero) when all the
%network has only one label.
The entropy $E$ reaches the maximum only for the flat distribution, where each
node knows only itself, and reaches a minimum (zero) when all nodes know the
same label (i.e. all state vectors are the same and contain just one element
different from zero).
If the population is evenly distributed
among $n$ clusters, the entropy is $E=\log(n)$.  
The value of Entropy $E(t)$ allows us to
discover the structure of the network, where different levels of the
hierarchical structure are identified by plateaus as shown in \figurename~\ref{fig:g2} (d).

%XXX

Now the question is: is it possible to design the algorithm independently from
the parameters?
%Here, we explore a ``cognitive algorithm'' in order to solve
%this task. We first  define the concept of  \emph{cognitive dissonance}
%between adjacent agents as described in the following equation:

\addedstuff{In order to solve this task, we explore a ``cognitive algorithm''.
%Here, we explore a ``cognitive algorithm'' in order to solve this
%task.
We first define the concept of cognitive dissonance between adjacent
agents. Cognitive dissonance has been defined within the field of social
psychology from Leon Festinger \cite{citeulike:235567}, in order to explain
the natural tendency of
people to reduce conflicting cognitions creating a consistent belief system,
or alternatively by reducing the importance of any source of dissonant
elements (e.g. sometimes friends or neighbors). The theory shows a good
predictive power, shedding light on otherwise apparently irrational or
destructive behavior, and can be reduced in our work as described in the
following equation:}
\begin{equation}
D_{ij} = \frac{|S_{i}-S_{j}|}{2},
\end{equation}
that is the difference between the absolute values of the state vectors between
$i$ and $j$. Then, we define the local entropy for each node at time $t$,
considering the state matrix $S$:
\begin{equation}
E_{i}^t=-\sum S_i^t\log{S_i^t}.
\end{equation}

In \figurename~\ref{fig:g2} (d), we show the global entropy of information
of the network during the time. The three plateaus correspond to three
different levels: if we evaluate the first derivative of the entropy we
can identify three peaks, while in the second derivative, we observe three
changes of sign. For this reason, we evaluate the first and the second
derivative of the local entropy for each node.
%In \figurename~\ref{fig:local_entropy}, we plotted the local entropy for a single node
%during the time at the end of the exploration phase. We can see how the node is able
%to discover three different plateaus corresponding to three levels of the
%hierarchical network showed in \figurename~\ref{fig:g2} (a).
Analogously for the entropy defined above, it is possible to introduce the
concept of local entropy for each node in order to study the local view of
agents. Similar as we can observe in \figurename~\ref{fig:g2} (d), it is possible
to detect different plateaus corresponding to the different network sub-clusters
that the single node discovers during time. We observed
that we can use a fixed value of the parameter $m$, while we have to change the
value of $\alpha$ in order to find the community and in particular the hubs that
labels each community. For this reason, we simulate an exploration phase of the
network several times in which the nodes save their state vector $S_i^t$, in a
\emph{temporary memory box}, when they observe a change in sign of the second
derivative. If the following condition is satisfied
\begin{equation}
\label{second_derivative}
sign\left(\frac{\delta^2 E_i^{t-1}}{\delta t^2}\right) \neq sign\left(\frac{\delta^2 E_i^{t}}{\delta t^2}\right),
\end{equation}
the state vector $S_i^t$ is stored into the temporary long term memory together
with the value of the first derivative of the local entropy and the entropy. 
When a node meets an impasse (e.g. its state vector entropy and its cognitive
dissonance do not evolve anymore) its $\alpha$ is changed by the following
mechanism, if
\begin{equation}
\left|\frac{E_i^{t-1}+D_i^{t-1}}{K_i}\right|-\left|\frac{E_i^{t}+D_i^{t}}{K_i}\right|< \epsilon,
\end{equation}
where $K_i$ is the connectivity degree of the node $i$. Then, a counter $\tau$ is increased by $1$
($\tau_i \leftarrow \tau_i +1$), and if $\tau_i$ becomes greater than a given
threshold (say $\tau^*$), the parameter $\alpha$ is updated in the following
way:
\begin{equation}
\alpha_i = 1.5  |\eta \sigma |+1,
\end{equation}
where $\eta$ is a random Gaussian variable with mean 1 and standard deviation
$\sigma$.  
\begin{figure*}[t!]
\centering
\subfigure[]
{\includegraphics[height=4cm]{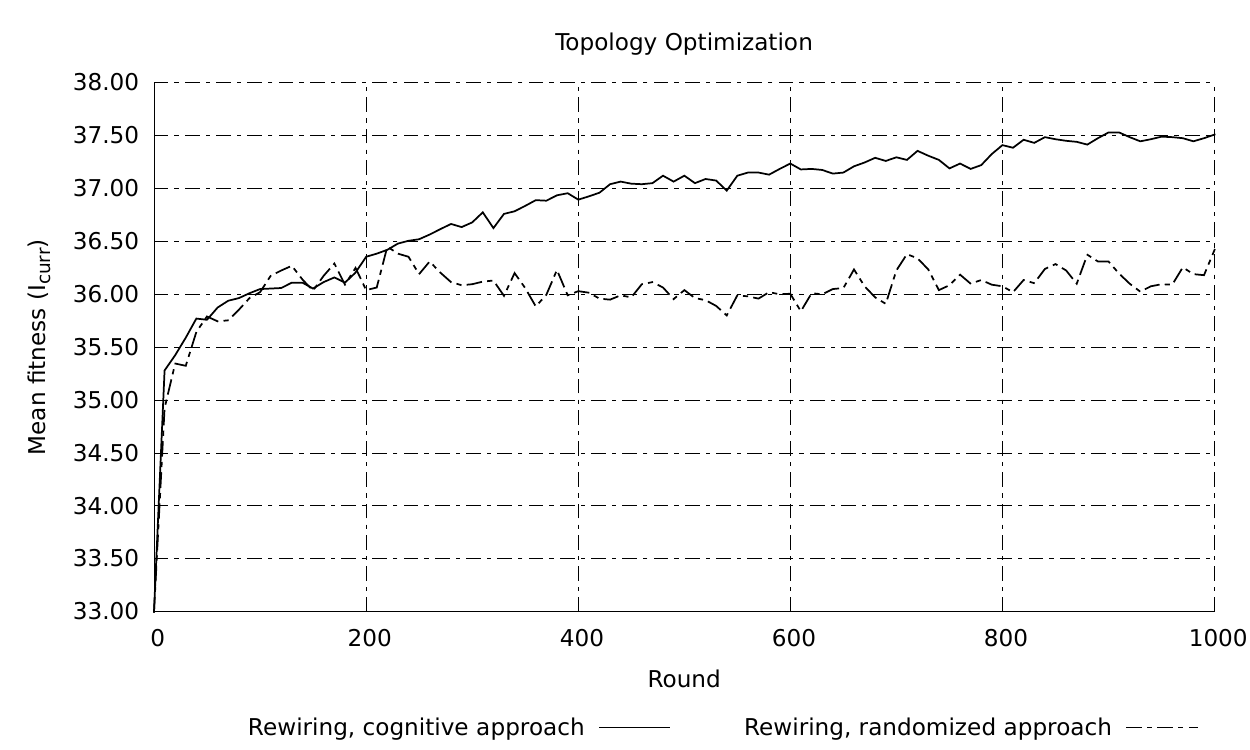}}
\hspace{.1mm}
\subfigure[]
{\includegraphics[height=4cm]{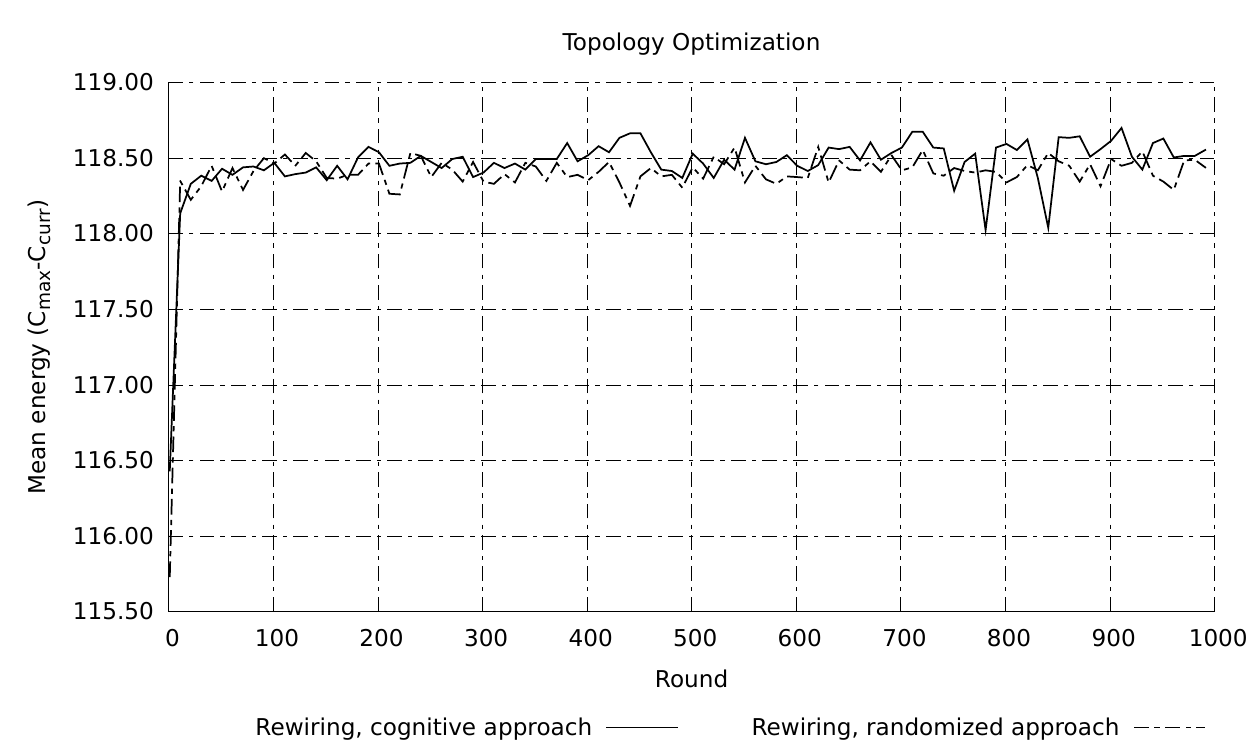}}
\caption{\label{fig:g3} (a) Fitness function that shows the mean number of
retrieved items per node.
%The network consisted of $N=200$
%nodes, energy per node was $E_{max} = 120$, number unique objects distributed
%across the network were $max_{obj} = 50$, the number of unique objects a node
%should retrieve with the given energy constraint was $n_{obj} = 45$.
The fitness of the randomized algorithm is
represented as the dashed line, our approach as the solid line. (b) Mean
energy usage per node of the randomized algorithm (dashed) and our approach
(solid).}
\end{figure*}
After a typical period of a fixed length ($\Delta T$), the process is stopped
for all nodes and a node's long term memory is updated with a new sample respectively experience. 
The long term memory is characterized by a bound threshold $B^1$ (here $B^1=5$)
in order to mimic the ecological limits of such cognitive functions (i.e. bounded
rationality).  After the node has saved its state vector when the sign
of its second entropy derivative changed (eq. \ref{second_derivative}), it
proceeds in structuring its long term memory. First, its first derivatives are decreasingly sorted, and then the first $B^1$
time positions are recorded. Later, such $B^1$ element vectors are descendingly
sorted with respect to the entropy. Finally, using the time positions,
the correspondent state vectors is analyzed and larger elements for each
state vector are assumed as potential hubs and therefore stored into the long term
memory. At this stage, the long term memory of each node is composed by a list
of $B^1$ sets of potential hubs, ordered following the procedure from the more
local to the more global one. Moreover, the long term memory is bounded by
another threshold ($B^2$), which represents the long term memory buffer,
i.e. the maximum number of the $B^1$ sets it can consider/contain, so that
the long term memory is represented by a $(B^1,B^2)$ matrix. Finally, each
node summarizes its knowledge of the network building a \textit{hub list}
obtained by analyzing the frequency in which each hub appears within the long
term memory, which is subsequently ordered from the most represented (i.e.
the hub with a larger frequency) to the least represented one. The knowledge
of the network (i.e. the hub list) is used by weak nodes in order to increase
the fitness.

\addedstuff{The nodes' fitness is computed in a general and conservative way following the ratio presented in section II. In the first scenario, the nodes are sorted with respect to the number of objects they collected trough their neighbors, while in the second scenario, they are sorted with respect to the amount of energy they spent to collect the maximum number of items. After this phase, the last $9\%$ of the nodes (e.g. the weakest nodes) are chosen for the cognitive rewiring, and in addition $3\%$ of the nodes are chosen for a random rewiring.}

Whenever a node does not have a ``sufficient''
fitness, it eliminates a portion of unnecessary links (i.e. those links which
point to nodes detected as non-hubs, in this work just $1$ link) and proceeds to try to establish
new connections using the hub list it has. Starting
from the most relevant hub (i.e. the first from the list) and continuing towards
the last one, the rewiring node tries to build new links. Finally, if no hubs
have available links, because they have reached the maximum number of connections,
the rewiring node adopts a random strategy and establishes a link towards the first available node it finds.
%XXX
%\begin{figure}[htb]
%\centering
%{\includegraphics[height=4cm]{entropy3.pdf}}
%\caption{\label{fig:local_entropy} Local entropy of information regarding a single node}
%\end{figure}

%In this last part, the ``rewiring'' phase, nodes have to find an optimal
%way to cut old connections and to wire new ones in order to minimize their
%energy and to maximize their ``fitness'', defined as function of the
%collected items. Bla bla....
%
%
%\begin{figure*}[t!]
%\centering
%{\includegraphics[width=10.5cm]{peaks.pdf}}
%\caption{\label{fig:g5}  ........}
%\end{figure*}

%\subsection{Rewiring phase}
%\label{sec:mod3}
%ToDo

\section{Evaluation}

We compared our model from section \ref{sec:model} with a randomized algorithm.
For comparability reasons between the algorithms, they are kept similar, apart
that the randomized algorithm is memoryless and therefore 
nodes have no knowledge about its surrounding and potential hubs they might
connect to. Consequently, the randomized algorithm selects the nodes that have
to rewire using the same method as the cognitive algorithm does; but where the
cognitive algorithm prefers to connect to a hub, the randomized one chooses a
random node.
\begin{figure}[b!]
\centering
{\includegraphics[height=4cm]{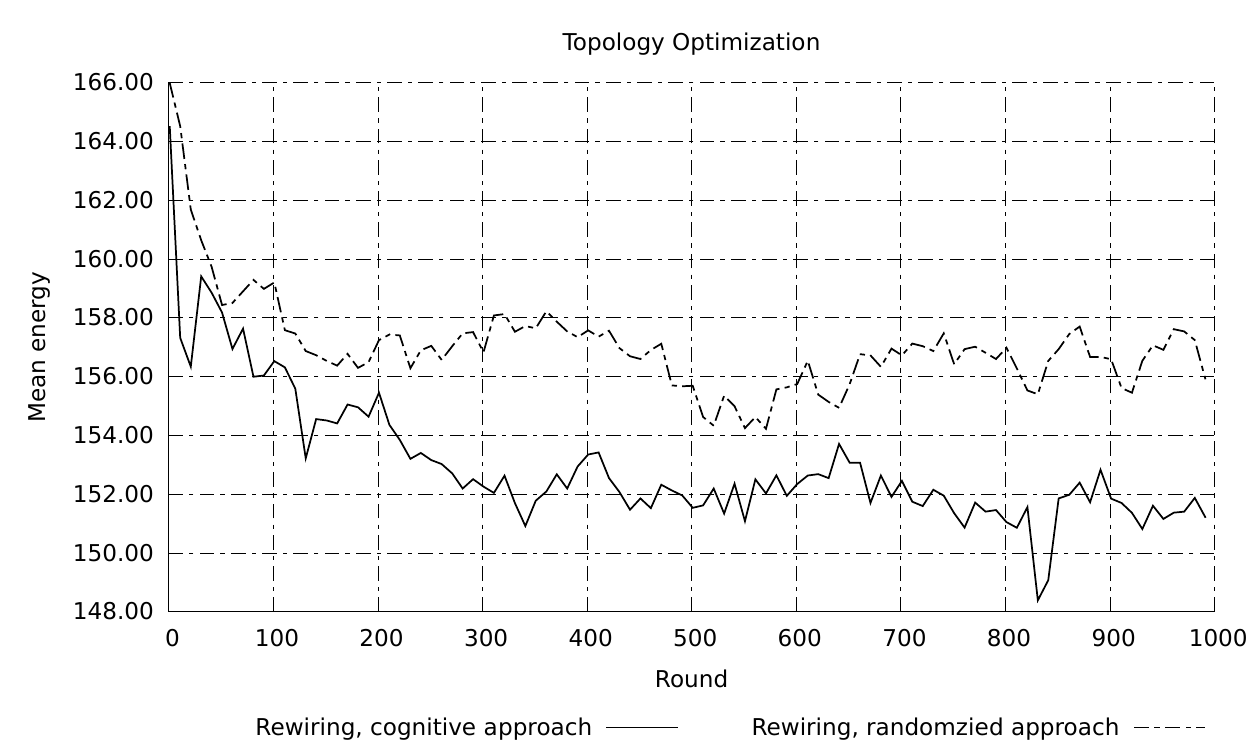}}
\caption{\label{fig:g4} Energy minimization approach: 
mean energy usage per node of the randomized algorithm (dashed) and our approach (solid).
%in this scenario, the
%randomized algorithm (dashed) and our approach (solid) had no energy constraints
%so that all items were retrieved, but it was rather the energy usage that has
%been minimized.
}
\end{figure}

For the evaluation, we used the two scenarios described in section \ref{sec:sce} in
order to test our algorithms.
%They differ in
%the measurement of the quality of the created network. In the first one, each
%node has a \textit{fixed} amount of energy, that can be spend for hops. The goal is to retrieve as much items as
%possible without a violation of the energy restrictions. This refers to the
%fitness function $f_1$ from section \ref{sec:sce}.
%In the second one, each node retrieves \textit{all} items in every round.
%The objective is to minimize the average used energy that correlates as
%described before to the amount of hops needed by each node. This refers to the
%fitness function $f_2$ in the scenario description (section \ref{sec:sce}).
The initial network topology consists of $N = 200$ with a mean connectivity
per node of $4$. A total of $I = 50$ unique items is distributed
among the nodes, where each node needs to retrieve $I_{max} = 45$ objects from
its neighbors.
We used this setting in order to analyze a network on a larger scale.
%We used this setting, because our goal was to use a network as
%large as possible regarding the given computation time.
Further, we also tested the algorithm for smaller networks, and the results
imply a similar behaviour as presented here.
%For them, 
We run the simulation $50$ times on our Matlab 
cluster with different random seeds. \figurename~\ref{fig:g3} shows an initial
result for the first scenario and \figurename~\ref{fig:g4} for the second. Both
figures show values of the median run regarding final results of fitness and energy.

In the first scenario, the number of retrievable items shall be maximized.
Therefore, the ``weakest'' nodes are determined by the sum of collected items.
%calculation of the number
%of collectable items. 
In \figurename~\ref{fig:g3}, it is shown that both approaches
improve the initial topology significantly at the beginning. After having
reached a plateau of $36$ items, the randomized approach begins to oscillate,
whereas the cognitive approach can exploit its knowledge of potential hubs
and steadily micro-optimizes the topology up to more than a \textit{mean} of
$1.1$ items by not having significant differences in their energy usage. We
can also observe that the cognitive approach is less prone to oscillations.

The second scenario shown in \figurename~\ref{fig:g4} shows the energy dynamics of
both approaches. Each node has unlimited energy available, so that it is able
to retrieve all necessary $45$ items. The weakest nodes are now defined as nodes
who consume the most energy of all. Hence, those are candidates for rewiring
in order to minimize the system's energy. The behaviours of both approaches are
quite similar as in the fitness optimization from \figurename~\ref{fig:g3}. The
initial topology improvement significantly reduces the energy consumption of
the system. However, oscillation effects occur more often than in the first
scenario. Our cognitive approach reduces the \textit{mean} energy consumption
of the nodes of more than $4.1$ hops per node compared to the randomized
algorithm.

\section{Conclusion}
In this work, we described how we optimize a topology by the means of a
cognitive-inspired algorithm. The resulting online optimization problem was tackled
with a cognitive model that enables a node to be self-aware about its
surrounding community and eventually to detect and distinguish between hubs and
non-hubs. This knowledge was exploited by a node to gain a more effective
rewiring to other nodes than by random selection.
We showed the effectiveness of our approach in two scenarios, in each comparing
the achieved results to a randomized algorithm using the same network
conditions.
In the first scenario, the goal was to find a topology in which a maximum number of unique items
can be retrieved for the system under a given energy constraint that was spent
for ``hopping''. In the second one, we removed the energy constraint, so that nodes had
enough energy for retrieving all items in each round, with the focus on
decreasing the system's overall energy.
In both scenarios, the cognitive-inspired algorithm performed significantly
better than the random one.

%the nodes should use as little hops as possible to retrieve all different items.

%% OK!
Despite the fact that the algorithm uses global information for the selection of rewiring
nodes, the approach shows first steps towards a pure self-organizing
network since only local information is used for the hub detection.
Overall, we showed first steps that information generated by a
cognitive-inspired algorithm can be exploited in order to optimize network
topologies. 
As future work, we plan to (i) deploy the algorithm on a wide range of
\textit{large scale} network topologies, (ii) localize the decision making
of a node when to rewire or not, and (iii) further elaborate the used
scenario by introducing more dynamics into items and nodes. \addedstuff{We think that 
our algorithm is generic enough that it could also be used as a foundation
in a wide area of applications beyond the scenario proposed here.}

%it could also be applied as groundwork in different
%scenarios than presented.}

%In future work the results have to be enlarged to a wider range of starting
%networks. At first the number of nodes and the mean connectivity will be varied.
%In a second step the starting topology might be changed, for instance to a small
%world network.
%
%Another point for the improvement of the algorithms is the focus on local
%information only. In the algorithm shown here the nodes decide on the basis of
%the fitness of the hole network weather or not they rewire themselves. In future
%versions of the algorithm this decision will be localized.
%
%The third important goal is to proof the direct applicability to one of the
%named examples in the introduction. Our goal in this work was to create an easy
%generic scenario in which the usefulness of the approach can be shown. After
%this is done we want to be more concrete in our application.

\section*{Acknowledgment}
We would like to thank the organizers of the Awareness Summer School 2012
where we started our work on this topic. This research has received funding from
the European Union 7th Framework Programme under grant agreement n$^\circ$ 257906
and n$^\circ$ 257756.

\bibliography{sample-paper}
\bibliographystyle{plainnat}

\end{document}